# On the origin of diverse time scales in the protein hydration layer solvation dynamics: A molecular dynamics simulation study

**Sayantan Mondal[1], Saumyak Mukherjee[2]** and **Biman Bagchi***

Solid State and Structural Chemistry Unit
Indian Institute of Science, Bangalore, India
E-mail: *profbiman@gmail.com

## Abstract

*In order to inquire the microscopic origin of observed multiple time scales in solvation dynamics we carry out several computer experiments. We perform atomistic molecular dynamics simulations on three protein-water systems namely, Lysozyme, Myoglobin and sweet protein Monellin. In these experiments we mutate the charges of the neighbouring amino acid side chains of certain natural probes (Tryptophan) and also freeze the side chain motions. In order to distinguish between different contributions, we decompose the total solvation energy response in terms of various components present in the system. This allows us to capture the interplay among different self and cross-energy correlation terms. Freezing the protein motions removes the slowest component that results from side chain fluctuations, but a part of slowness remains. This leads to the conclusion that the slow component in the ~20-80 ps range arises from slow water molecules present in the hydration layer. While the more than 100 ps component may arise from various sources namely, adjacent charges in amino acid side chains, the water molecules that are hydrogen bonded to them and a dynamically coupled motion between side chain and water. The charges, in addition, enforce a structural ordering of nearby water molecules and helps to form local long-lived hydrogen bonded network. Further separation of the spatial and temporal responses in solvation dynamics reveals different roles of hydration and bulk water. We find that the hydration layer water molecules are largely responsible for the slow component whereas the initial ultrafast decay arise predominantly (~80%) due to the bulk. This agrees with earlier theoretical observations. We also attempt to rationalise our results with the help of a molecular hydrodynamic theory that was developed using classical time dependent density functional theory in a semi quantitative manner.*

---

[1] E-mail: sayantan0510@gmail.com
[2] E-mail: mukherjee.saumyak50@gmail.com
* corresponding author

# I. INTRODUCTION

Water molecules that surround a protein play a major role in providing stability to its native structure. Biological activities of proteins are also dependent on the dynamics of protein hydration layer (hereafter referred to as PHL).[1-17] Because of its immense importance in biology, PHL has remained a subject of substantial interest.[4,15-28] With the help of various theoretical and experimental methodologies, new insights into the nature of PHL are emerging.[29,30]

Surprisingly, however, different experimental and simulation studies seem to have revealed rather divergent results regarding the slow timescale of relaxation inside PHL.[2] Our present study aims to unveil the microscopic details and to establish the origin of slow timescale in solvation dynamics. Thus we attempt to resolve some of the long standing debates in this field.

*(i) What are the prevailing factors responsible for the slow solvation in PHL: water, protein or a coupled motion?*

*(ii) Why do different experiments give such different results?*

*(iii) What are the precise origins of the diverse time scales that seem to range from tens of fs to hundreds of ps?*

An early estimation of PHL came from the elongated rotational time constants of water molecules measured by nuclear magnetic resonance (NMR) experiments in several aqueous proteins. This eventually directed to the concept of the 'iceberg model'[31,32] that was later dispelled by Wüthrich[28,33]. Even earlier than the reported NMR results, Pethig, Grant and others[18] found three discrete timescales in dielectric relaxation namely, ~10 ps, ~10 ns and ~40 ps (the δ-dispersion attributed to PHL) that were later verified by Mashimo[20,34].

Much later, this problem was revisited by Halle *et al.* using an improved NMR technique widely known as magnetic relaxation dispersion (MRD)[26,35]. Halle *et al.* suggested that any relaxation process is slowed down maximum by a factor of ~2-4. More recently, Kubarych *et al.* have used 2D-IR spectroscopy to find out the orientational slowness of PHL that corroborates well with the MRD data[36]. On the contrary, several time dependent fluorescence Stokes shift (TDFSS) and computer simulation studies were carried out by Zewail, Bhattacharyya, Maroncelli, Fleming and others[7,8,12,37-40] that reported observation of a component in the long time that was substantially slower than what was observed in the bulk.

One aspect that is often ignored in the discussions of PHL dynamics is the sensitivity of the dynamics to different experimental probes placed at different locations. That is the dynamics is highly inhomogeneous. Negation of this can lead to incomplete understanding. For example, NMR studies provide single particle dynamics but averaged over a large number of molecules[35]. On the other hand, dielectric relaxation sums over contribution from all the molecules and hence a collective measure[6,41].

In a series of studies, Zewail *et al.* observed the presence of a slow component (one less than 1ps and another in 20-40 ps range) in solvation of tryptophan bound to *Subtilisin Carlsberg* and *Monellin* using TDFSS. But they were unable to detect the ultrafast component because of limited resolution.[7,42] Bhattacharyya *et al.* also obtained timescales in the order of few hundred ps to a few ns.[8,39,40] Much later, Zhong *et al.* employed site directed mutations on sperm whale myoglobin which led to a spectrum of relaxation timescales (1-8 ps and 20-200 ps)[43] that are present at different sites of the same protein. The same group has also studied the effect of charges on the timescales using alanine scan method at different sites.[44]

TDFSS experiments measures the time dependent frequency, $\nu(t)$, of a fluorescence probe in order to construct a non-equilibrium stokes shift response function (**Eq.[1]**).[1,38,45-48]

$$S(t) = \frac{\nu(t) - \nu(\infty)}{\nu(0) - \nu(\infty)} = \frac{E_{solv}(t) - E_{solv}(\infty)}{E_{solv}(0) - E_{solv}(\infty)} \quad [1]$$

Where, $E_{solv}(t)$ is the time-dependent energy response as measured by the probe at time '$t$' (**Eq.[2]**).

$$E_{solv}(t) = -\tfrac{1}{2} \int d\underline{r}\, \boldsymbol{E}_0(\underline{r}) . \boldsymbol{P}(\underline{r}, t) \quad [2]$$

Here, $\boldsymbol{E}_0(\underline{r})$ is the position dependent bare electric field of the polar solute (Tryptophan in our case). $\boldsymbol{P}(\underline{r},t)$ is the position and time dependent polarisation[7]. Solvation of an ion in a dipolar liquid is faster compared to dielectric relaxation. Whereas, solvation time constant for a dipole ($\tau_L^d$) is slightly higher than that of an ion ($\tau_L$) (**Eq.[3]**).[37]

$$\tau_L = \left(\frac{\varepsilon_\infty}{\varepsilon_0}\right)\tau_D$$

$$\tau_L^d = \left(\frac{2\varepsilon_\infty + \varepsilon_c}{2\varepsilon_0 + \varepsilon_c}\right)\tau_D \quad [3]$$

Here, $\tau_D$ is the Debye relaxation time; $\varepsilon_c$ is the dielectric constant of the molecular cavity; $\varepsilon_0$ and $\varepsilon_\infty$ are respectively the static and infinite frequency dielectric constants of the solvent. The complex solvation in the bulk gets even more complicated for hydration water. In addition to the multitude of timescales there is a dominant role of coupling between the motions of side-chain and water, which still deserves proper quantification. According to a "*unified model for protein dynamics*" proposed by Frauenfelder *et al.* there is 'slaving' of small scale and large scale protein motions by hydration and bulk water molecules respectively.[5,49] Also the determination of a local an effective dielectric constant is non-trivial and approximate.[29,30]

MD simulations reveal that the freezing the protein motions results in a faster solvation at W7 site of apomyoglobin.[50] Zhong *et al.* showed that charged/polar residues as well as the side-chain fluctuations play a major role in slowness.[15,44,50] It is straightforward to understand the reason of fastness on freezing side chain motion because a slow component is removed from time dependence of solvation energy. However, the role of water molecules inside the PHL in slowing down the dynamics and their relative contribution is not properly addressed.

Hence, to seek the answers to those questions that are raised in the beginning, we choose several intrinsic natural tryptophan (W) probes in (i) Myoglobin (W7) (ii) Lysozyme (W123) and (iii) sweet protein Monellin (W3). Mutation of the neighbourhood of natural probes and decomposition of the solvation energy into various components reveal information on how they can affect the local structure and dynamics. Our present work provides a logical and semi-quantitative explanation of the timescales of solvation dynamics at different sites of the protein and also the reason for their existence.

Another important aspect that is often overlooked is the spatial dependence of the temporal response. Both TDFSS and dielectric relaxation experiments measure response which is sum of the response from water molecules in the first layer and the outer layers. A molecular hydrodynamic theory predicts dynamic response to depend on the length scale of the process. Thus, first layer response is predicted to be slower than that of the bulk. Our results, presented in the subsequent sections below, are also consistent with this prediction.

The organisation of the rest of the paper is as follows: In the next section (**Section II**), we discuss the details of computer experiments. In **Section III**, we report and explain the results obtained by analysing the MD trajectories. This section is divided into several sub-sections. In subsequent sections (**Section IV** and **V**) we rationalise the obtained numerical data with the help of a time dependent density functional theory (TDDFT) based molecular hydrodynamic theory (MHT) description. In **Section VI**, we end with the importance of these findings on solvation dynamics, and also with some general conclusions.

## II. COMPUTER EXPERIMENTS AND ANALYSES METHODS

We perform the following two kinds of computer experiments- (a) Mutation of the neighbourhood amino acid side chain charges (see *Supporting Information* **S1** for details) that reside within 7-8 Å of the probe and (b) Freezing the whole protein's natural motions. The first set of experiments allows us to investigate the role of charges in the surroundings of a natural probe on solvation dynamics. Whereas, the second set of experiments allows us to remove the effect of the innate amino acid side chain fluctuations.

So there can be following four such circumstances. (i) A non-zero net charge around the probe and protein is mobile, (ii) No net charge around the probe but protein is mobile, (iii) There is a net charge around the probe but protein is frozen and (iv) No net charge around the probe and the protein is also frozen. Solvation time correlation functions (TCFs) are then calculated for wild type and mutated proteins in both mobile and frozen state. The natural surroundings (i.e., wild type) of the chosen probes are noted down in **Table 1**.

**Table 1.** Surroundings of the selected natural probe tryptophan (W) in Lysozyme, Myoglobin and Monellin. The standard single letter amino acid codes are used in this table. [E=Glutamate, K=Lysine, D=Aspartate and R=Arginine]

| Probe | Protein | Residues in the neighbourhood | Surroundings (wild type) | Surroundings (mutant) |
|---|---|---|---|---|
| **W7** | Myoglobin | E4, E6, K77, K78, K79, K133 | Charged | Charges removed |
| **W123** | Lysozyme | R5, K33, R114, K116, D119, R125 | Charged | Charges removed |
| **W3** | Monellin | E2, E4, K44, E59, D74 | Charged | Charges removed |

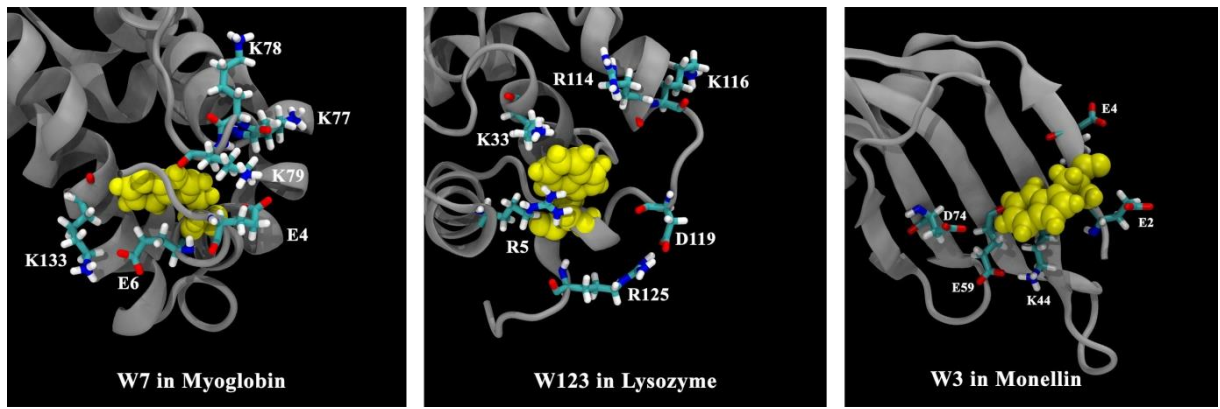

**Figure 1.** Snapshots from MD simulations (initial frame at t=0) that shows the natural neighbourhood of selected natural probes, W7 in Myoglobin (left), W123 in Lysozyme (middle) and W3 in sweet protein Monellin (right). All the probes are embedded in a polar environment provided by the nearby amino acid side chain residues in their wild form. Tryptophan residues (used as a natural probe) are shown in yellow. The figures are prepared using VMD.[51]

The subsequent analyses are performed in the following way. First, we observe the local structural ordering of water molecules around a natural probe with/without the presence of neighbourhood charges from radial distribution function (RDF) between relevant side-chain atoms and water.

Next we compute the solvation energy time correlation functions. Although S(t) [in **Eq.1**] is a non-equilibrium response function, under the assumption of linear response theory[52], we can calculate S(t) from the equilibrium energy time correlation function C(t) (**Eq.[4]**).[38,46,47,53-55]

$$C(t) = \frac{<\delta E_{solv}(0)\delta E_{solv}(t)>_{gr}}{<\delta E_{solv}(0)^2>_{gr}} \quad [4]$$

Here, $\delta E_{solv}(t)$ is the fluctuation given by; $\delta E_{solv}(t) = E_{solv}(t) - <E>$. The subscript '$gr$' indicates averaging over ground state only.

For further analysis, we decompose the time dependent solvation energy $E_{solv}(t)$ into various contributions that arises from several sub-ensembles, namely side-chain (SC), protein core (Core), water (Wat) and ions (Ion) as depicted in **Eq.[5]**.[55,56]

$$E_{solv}(t) = E_{SC}(t) + E_{Core}(t) + E_{Wat}(t) + E_{Ion}(t) \quad [5]$$

We find contributions from terms involving ions are negligible compared to others. Hence, we can express total solvation time correlation function as a summation of three self and six cross-correlation terms (**Eq.[6]**).

$$S(t) = \sum_{\alpha} S_{\alpha\alpha}(t) + \sum_{\alpha}\sum_{\beta} S_{\alpha\beta}(t) \quad [6]$$

Where, $\alpha$ and $\beta$ stand for different components. The normalised total solvation energy correlation plots are fitted to a multi-exponential form along with a Gaussian component ($\tau_g$) which provides information regarding the ultrafast dynamics (**Eq.[7]**).

$$S(t) = a_g e^{-\left(t/\tau_g\right)^2} + \sum_{i=1}^{n} a_i e^{-\left(t/\tau_i\right)}; n = 2 \quad [7]$$

$\tau_1$ and $\tau_2$ are another two timescales (intermediate and slow relaxation). It is justifiable on the ground that the ultrafast component is noticeably faster that the exponential ones and it carries ~100% weightage at t=0. The average time constants ($<\tau>$) are calculated by integrating S(t) with respect to time (0 to $\infty$)[48].

In order to precisely identify the prevailing factor(s) in slow and ultrafast part of dipolar solvation dynamics we further dissect the water contribution into two parts- (i) contribution arising due to water molecules inside ~1nm which characterise PHL and (ii) contribution arising due to the water molecules in the outer layer.

## III. RESULTS AND DISCUSSIONS:

### A. Enhancement of local structure of water in the presence of charges

We calculate the pair correlation function (i.e., RDF) between the $N_\varepsilon$ of indole moiety of the natural probe and oxygen atoms of water molecules in order to investigate the local structural changes of water in presence/absence of neighbourhood charges. For W123 in Lysozyme, an extra peak around ~4.7Å is present while there are charges nearby in the wild type lysozyme (**Figure 2a**). Similarly for W7 in Myoglobin, an enhancement of the first and second peak height is observed while the probe is surrounded by charges (**Figure 2b**). However, in case of W3 in Monellin, the wild type and the mutant shows almost similar g(r) plots. This is mainly because W3 in Monellin is sufficiently exposed to water in both wild and mutant type. Nevertheless, structural enhancement, though not much, can be noticed and the presence of the third small peak while surrounded by charges suggests long range ordering (**Figure 2c**).

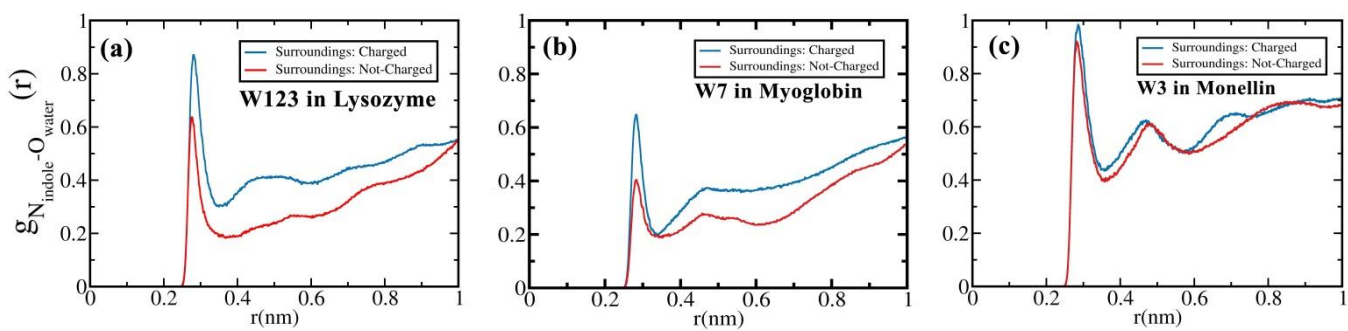

**Figure 2.** Radial distribution functions (RDF) for indole-nitrogen (of natural probe tryptophan) and water-oxygen pair. (a) RDF for W123 in Lysozyme; Wild type (blue) and neighbourhood mutated W123 (red). (b) RDF for W7 in Myoglobin; Wild type (blue) and neighbourhood mutated W7 (red). (c) RDF for W3 in sweet protein Monellin; Wild type (blue) and neighbourhood mutated W3 (red). This clearly shows how the presence of neighbourhood charges enforces local structural order in the nearby water molecules. [The colour codes are maintained such that *blue represents charged environment whereas red represents neutral environment*.]

**Figure 2** clearly shows the distortion of the water structure and certain rigidity that is not present in bulk water. These are the result of strong electrostatic interaction and long lived hydrogen bonds between the charges and the water molecules. Lowering of the peak height is also suggestive of the fact that a less number of water molecules are in the vicinity of the probe. This indeed affects the solvation timescales that we discuss in the subsequent sections.

### B. Solvation Dynamics of Tryptophan: Effect of adjacent charges and side-chain fluctuations

Solvation energy relaxation of natural probe tryptophan inside PHL shows ubiquitous bimodal nature along with a sub 100fs ultrafast component. However, the latter has been missed by many TDFSS experiments because of compromising resolution of the laser. W7 in the mutant myoglobin, where charged residues in the vicinity are made neutral, shows about a 2.5 fold decrease in average solvation relaxation compared to the wild type. Both the value and amplitude of the slowest timescale reduces (**Table 2**). W123 in mutated lysozyme also shows a faster relaxation when the charges are mutated to provide a neutral surroundings. The average solvation time constant also gets slightly reduced, although not to a great extent, on removal of the charges (**Table 2**). The slow timescale are comparable (120-130 ps) in both the cases. Surprisingly, W3 in monellin shows some anomaly. Because of the removal of charges, the initial decay was faster than that of the wild type as expected. But, at longer times (after 20 ps), the slowest timescale becomes larger although it possesses a smaller amplitude. Average timescales are however comparable. The reason of the anomaly may be speculated in terms of the immediate surroundings and also the solvent exposure of the probe. Each protein that is selected for our study falls in different classes (See ***Supporting Information* S2** for details).

Upon freezing the motion of the protein we can separate out the timescales that may arise solely because of the combined effect of water in the PHL and bulk. Solvation becomes faster. This was also observed by Singer *et al*.[12,50]

The slowest component above ~100ps disappears when the amino acid side chains are not fluctuating. This indicates the role of amino acid side chain fluctuations in a substantial retardation of solvation timescales.

But a slow component within ~20-80 ps is still present which is indeed a lot slower than bulk. In case of W7 in myoglobin it is *84.9 ps* (18%); for W123 of lysozyme it is *38.3ps* (7%) and for W3 in monellin it takes up a value of *23.5 ps* (4%). We attribute this to the slow orientational and translational motion of water inside PHL.[29] We further establish this fact in **Sec III.C.** Note that the trend (i.e. upon removal of charges/fluctuations) of the relaxation behaviour is not same for these three probes. Different sites responses differently that too varies from protein to protein.

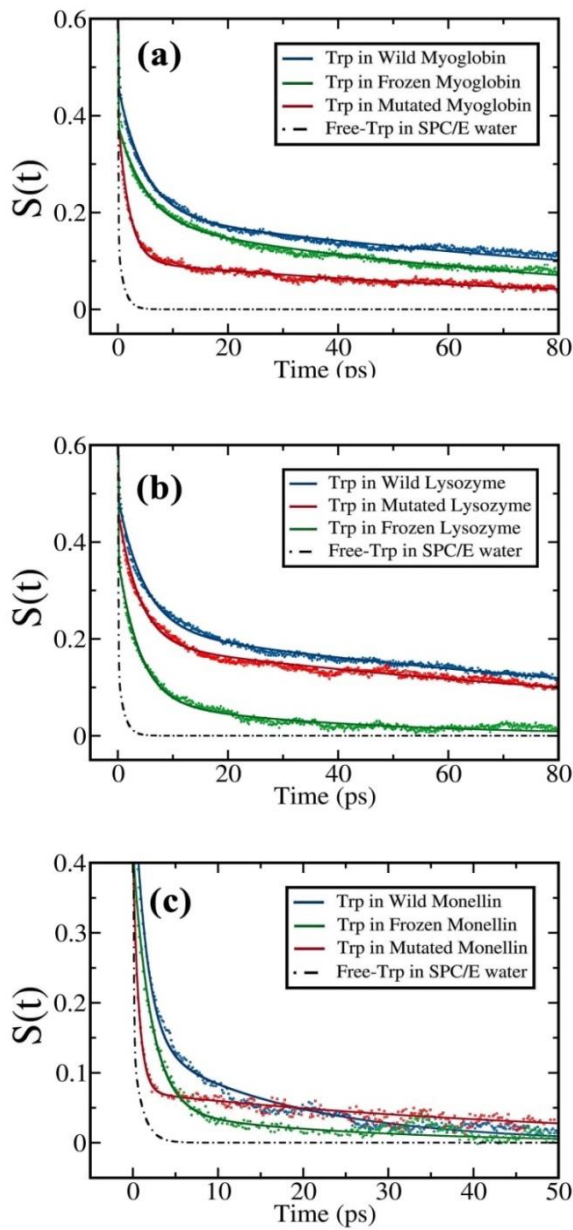

**Figure 3. Normalised total solvation energy time correlation function (TCF) for natural probes in Lysozyme, Myoglobin and Monellin. TCFs of probe with surrounding charges are shown in blue, without surrounding charges are shown in red and that of frozen conformations are shown in green. (a) W7 in myoglobin, (b) W123 in lysozyme and (c) W3 in monellin. Solvation energy relaxation for free tryptophan in SPC/E water is extremely fast as compared to that of protein surfaces (shown in black). The blue curve shows the slowest decay than the rest. That shows the role of surrounding charges and conformational fluctuation in retarding the solvation timescales. [*Initial ultrafast decay is ubiquitous and hence omitted from the plots*]**

**Table 2. Timescales (parameters after fitting to Eq.[7]) of solvation energy relaxation for two natural probes in Myoglobin, Lysozyme and Monellin, along with free tryptophan in SPC/E water model.**

| Probe | Protein | Surroundings | $a_g$ | $\tau_g$ (ps) | $a_1$ | $\tau_1$ (ps) | $a_2$ | $\tau_2$ (ps) | $<\tau>$ (ps) |
|---|---|---|---|---|---|---|---|---|---|
| W7 | Myoglobin (Mgb) | *Wild Type, Charged* | 0.54 | 0.092 | 0.26 | 4.80 | 0.20 | 120.6 | 25.4 |
| | | *Mutated, Charge removed* | 0.59 | 0.082 | 0.31 | 1.99 | 0.10 | 92.2 | 9.9 |
| | | *Wild type, Frozen* | 0.62 | 0.071 | 0.20 | 5.05 | 0.18 | 84.9 | 16.3 |
| W123 | Lysozyme (Lyso) | *Wild Type, Charged* | 0.51 | 0.094 | 0.27 | 4.83 | 0.22 | 130.4 | 30.0 |
| | | *Mutated, Charge removed* | 0.53 | 0.087 | 0.28 | 3.88 | 0.19 | 121.5 | 24.2 |
| | | *Wild type, Frozen* | 0.63 | 0.090 | 0.30 | 3.91 | 0.07 | 38.3 | 3.9 |
| W3 | Monellin (Mon) | *Wild Type, Charged* | 0.44 | 0.061 | 0.41 | 1.50 | 0.15 | 18.2 | 3.4 |
| | | *Mutated, Charge removed* | 0.59 | 0.037 | 0.34 | 0.69 | 0.07 | 52.3 | 3.9 |
| | | *Wild type, Frozen* | 0.58 | 0.067 | 0.37 | 2.12 | 0.05 | 23.5 | 2.0 |
| Free-Trp | --- | --- | 0.84 | 0.07 | 0.16 | 1.1 | --- | --- | 0.23 |

The above data indicate that the nearby charges play a major role in slowing down local dynamical responses. It is, again, primarily because of the long-lived hydrogen bonds those residues can form with surrounding water molecules which cannot orient rapidly (*see* **Sec III.E**).

**C. Decomposition of solvation energy into self and cross correlation terms**

According to **Eq.[5]**, we decompose solvation response, as measured by the natural probes, into several partial terms. Based on that, we calculate partial energy correlation functions for W123 and W7 in wild and mutant protein. There are some common features. Solvation of both the probes (**Figure 4a**) draws most of its contribution from Wat-Wat self-term. As these probes are exposed, water contribution is expectedly predominant.

Nevertheless, the SC-SC/Core-Core self-terms also adds a considerable slow component. Few cross terms are anti-correlated which indicates that increased contribution from one component results in a decrease of the other one. *The negative amplitude of such cross terms also helps in a faster relaxation and in dilution of slow timescales.*

In the case of W123; both SC-SC and Wat-Wat terms show substantial slow decays (**Figure 4**). Core-Core self-term and Core-Wat cross terms are negligible in amplitude. SC-Wat terms become anti-correlated and neutralise the huge slowness as well as amplitude arising from SC-SC and Wat-Wat terms. The faster decay for the mutant lysozyme is because of the presence of large amplitude slowly decaying anti-correlated SC-Wat cross terms, as compared to wild type. (**Figure 4**)

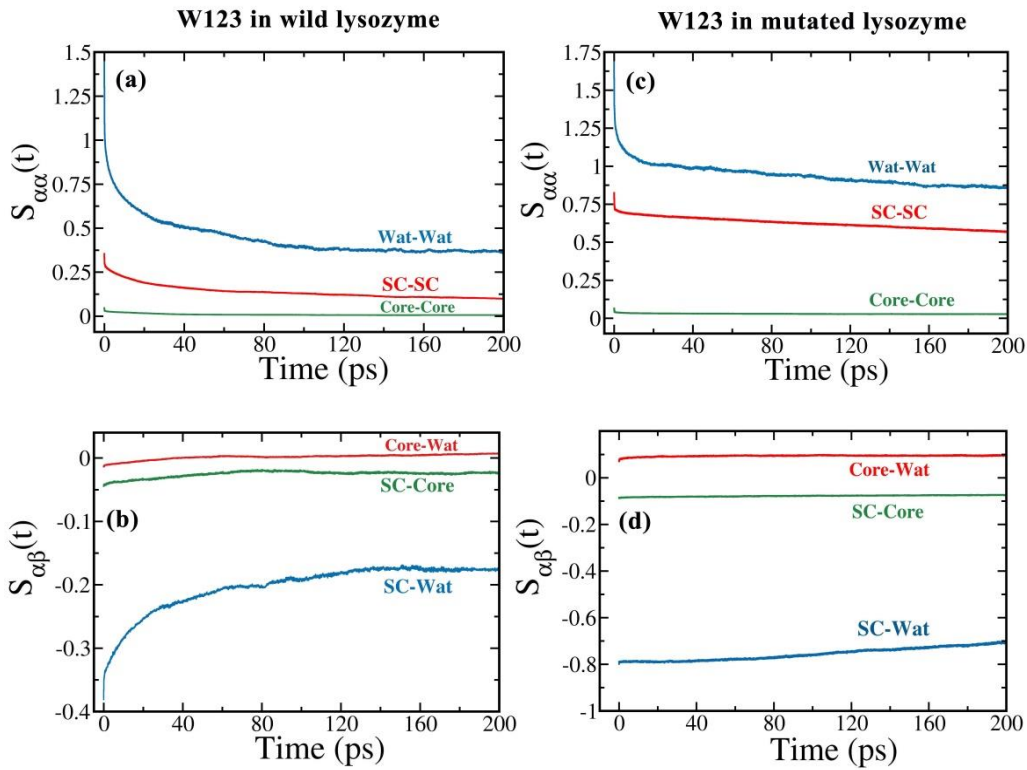

**Figure 4. Partial self and cross energy correlation terms, relative to the amplitude of total energy correlation, for W123 in wild type and its neighbourhood mutated protein. (a) self-terms in wild-type lysozyme. (b) cross-terms in wild-type lysozyme. (c) self-terms in mutated lysozyme. (d) cross-terms in mutated lysozyme. The presence of a noticeable slow component is in the Wat-Wat component. Anti-correlated cross terms are partly responsible for the faster decay of solvation energy relaxation.**

On the other hand, in the case of W7, in both wild and mutated myoglobin, interaction with protein backbone (core) plays an important role. When surrounded by charged side-chains, the SC-SC term has amplitude of about 0.2 that drops to nearly zero for the mutated protein. However in the wild type, SC-Wat terms are not anti-correlated (**Figure 5**). But, here as well, the water self-term plays a major role in the overall relaxation.

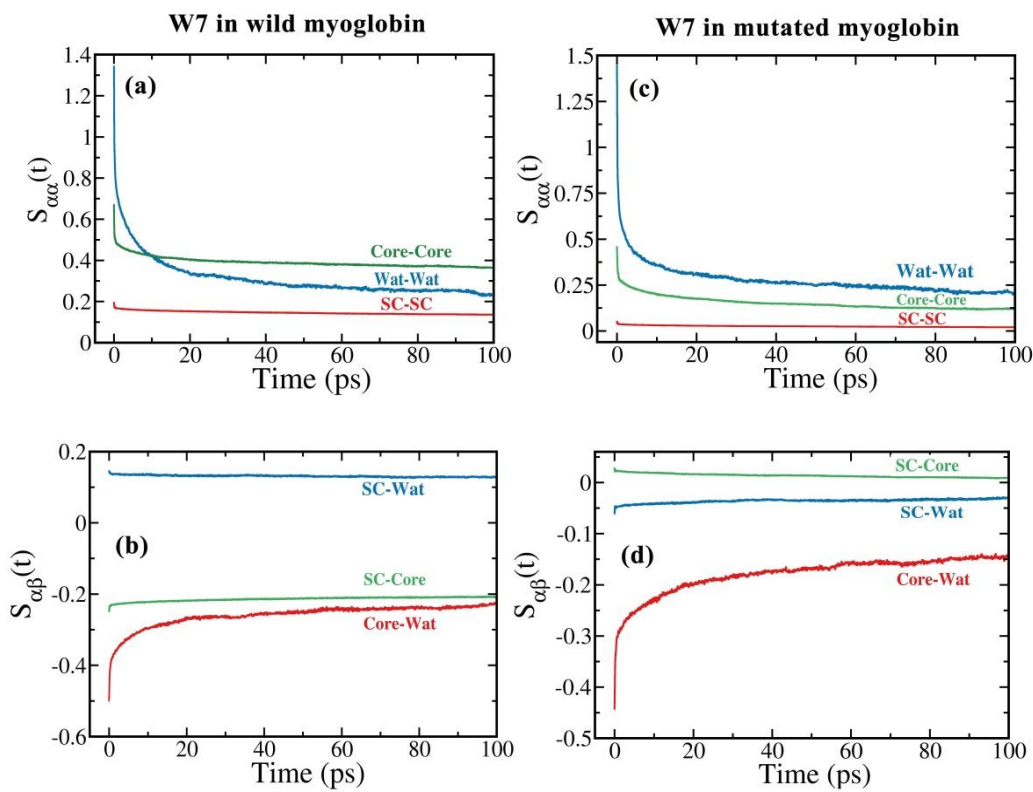

**Figure 5.** Partial self and cross energy correlation terms, relative to the amplitude of total energy correlation, for W7 in wild type and its neighbourhood mutated myoglobin. (a) self-terms in wild-type. (b) cross-terms in wild-type. (c) self-terms in mutant. (d) cross-terms in mutant. Here, unlike W123, the core is more strongly coupled with water compared to side-chain atoms. Remarkably, the SC-Wat cross term is not anti-correlated for wild type.

The partial correlation terms could explain **Figure 3**. Freezing results in a very fast decay for W123 but not so much for W7. This is because the neighbours (which contribute the most to the SC terms) of W123 are branchy (a lot of arginines) as compared to that of W7 (glutamets and lysines). As a result, the solvation energy relaxation becomes strongly dependent on the side-chain fluctuations in case of W123, but not to that extent in case of W7.

As it is clear from **Figure 4** and **Figure 5** that water plays a significant role in the overall solvation. The initial ultrafast decay is a contribution from water molecules in the outer layer. To look into the water term more closely we further separate the contribution from PHL and outer layer. We find that almost 80% of the ultrafast part (sub 100 fs) is mainly due to the outer layer. In fact, outer layer make no contribution to the slowly decaying part.

Therefore the slow solvation energy relaxation is a combined contribution from PHL and amino acid side chains. When the protein is mobile, the water molecules in PHL can alone contribute around 20-30% to the slow time component in the range of 200ps timescale that gets diluted in the presence of several anti-correlated cross terms. We find by freezing the protein that time constant of the slow decay remains around 120ps that approximately contributes 10%. When we do the mutation by removing the surrounding charges, we find a small increase in the rate. The absence of charges affects the slow component in PHL as the H-bond network is destroyed (see **Sec III.E**). The extra slowness arising from PHL in case of mobile protein is because of the presence of amino acid side chain motions that are coupled to the motion of water molecules inside PHL. **Figure 6** shows an exemplary plot for W7 of myoglobin.

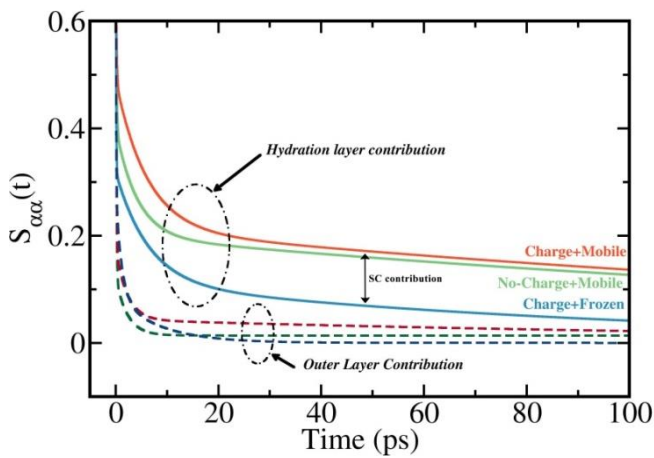

**Figure 6.** Normalised partial solvation time correlation functions that dissect the total water self-term into contributions from hydration layer and outer layer for W7 in myoglobin. This clearly shows the origin of fast relaxation is due to the ultrafast bulk modes and slowness arises from the hydration layer. Solid lines represent the contributions from PHL and dotted lines correspond to bulk counterparts. [*Colour codes are maintained such a way that the darker shades of the same colour represent the bulk contribution*]

This indicates that *the PHL water molecules can slow down solvation but not to that extent without the protein motions. This depicts the dynamically coupled motion between protein and hydration layer.*

**D. Solvation of probes in frozen proteins: Contribution from water**

We isolate the contribution arising solely from water by freezing the protein motions. So that, the self and cross terms involving 'SC' and 'Core' goes to zero and we are only left with water contribution. Of course, as discussed earlier, freezing results in a faster decay due to the absence of two main sources of slowness. These are (i) amino acid side chain motions and (ii) dynamical coupling between amino acid side chain and PHL. Still we find a slow component below 100 ps for each and every probe irrespective of charged/uncharged surroundings. In this case, there can be no other sources other than the slow water molecules in the PHL itself.

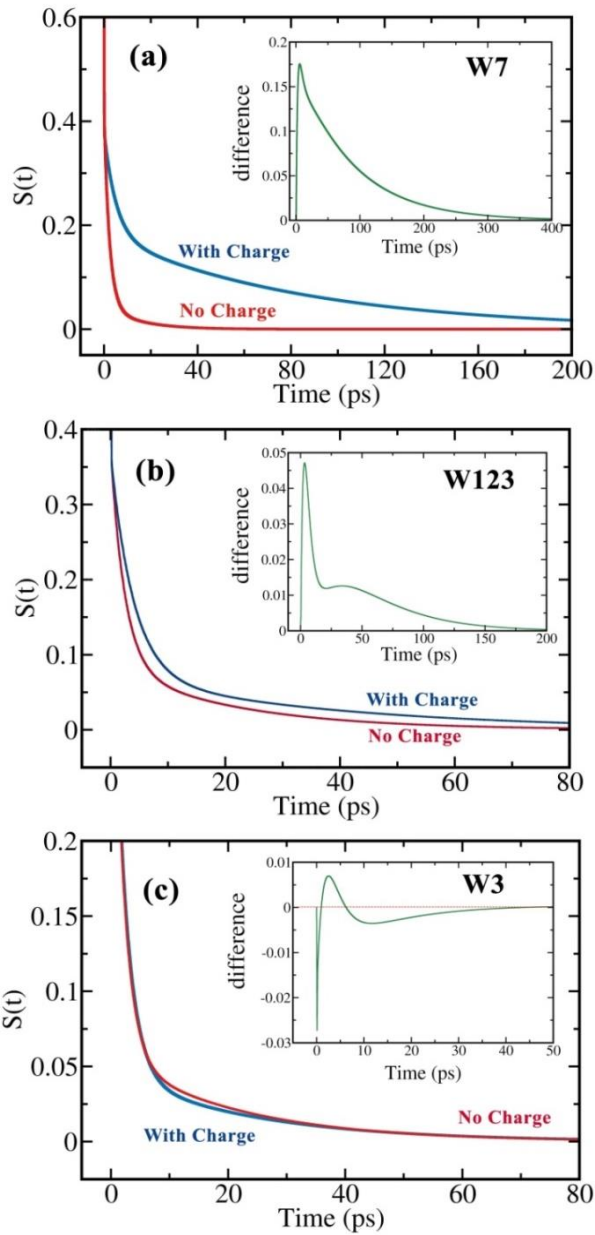

**Figure 7.** Comparisons of normalised total solvation time correlation functions (TCF) for tryptophan in frozen protein in case of polar (with charge) / non-polar (no charge) environment. (a) Normalised TCFs for W7 in frozen myoglobin. (b) Normalised TCFs for W123 in frozen lysozyme. (c) Normalised TCFs for W3 in frozen monellin. Insets contain the differences between two plots as sometimes it is hard to reconcile the relative slowness/fastness as they are too close. We note that a slow component remains irrespective of side chain fluctuations or neighbourhood charges.

**Table 3. Multi-exponential fitting parameters for the normalised solvation time correlation functions in case of frozen proteins** *(plots are shown in* **Figure 7***).* **The slow timescales even in the absence of side-chain motions are highlighted using boldfaces.**

| Probe | Type | Surroundings | $a_g$ | $\tau_g$ (ps) | $a_1$ | $\tau_1$ (ps) | $a_2$ | $\tau_2$ (ps) | $<\tau>$ (ps) |
|---|---|---|---|---|---|---|---|---|---|
| **W7** (Mgb) | Wild | Charged | 0.62 | 0.070 | 0.20 | 5.05 | 0.18 | **84.9** | 16.3 |
| | Mutated | Not Charge | 0.58 | 0.077 | 0.38 | 2.27 | 0.04 | **14.3** | 1.5 |
| **W123** (Lyso) | Wild | Charged | 0.63 | 0.090 | 0.30 | 3.91 | 0.07 | **38.3** | 3.9 |
| | Mutated | Not Charged | 0.62 | 0.067 | 0.29 | 2.44 | 0.09 | **21.3** | 2.7 |
| **W3** (Mon) | Wild | Charged | 0.58 | 0.067 | 0.37 | 2.12 | 0.05 | **23.5** | 2.0 |
| | Mutated | Not Charge | 0.56 | 0.070 | 0.38 | 1.85 | 0.06 | **21.4** | 2.0 |

**E. Hydrogen bond dynamics around W123 in lysozyme**

In order to establish the connection between the local H-bond dynamics and solvation timescales, we carry out hydrogen bond population analysis as well as lifetime studies between acidic hydrogens (i.e., H atoms connected to N-atoms of R5, R114, R125, K33 and K116) that surround W123 and water molecules. The well-known geometrical criteria (angle and distance) are employed as discussed by Chandler *et. al.*[57] to define H-bonds. The H-bond lifetime is then investigated using two types of correlation functions which are constructed of Heaviside step functions - (i) the intermittent H-bond correlation function ($\frac{<h(0)h(t)>}{<h(0)>}$) and (ii) Continuous H-bond time correlation function ($\frac{<h(0)H(t)>}{<h(0)>}$)[58,59]. Here h(t) adopts a value of '1' if the bond exists and '0' otherwise. On the contrary H(t) becomes continuously '0' if the bond breaks for once.

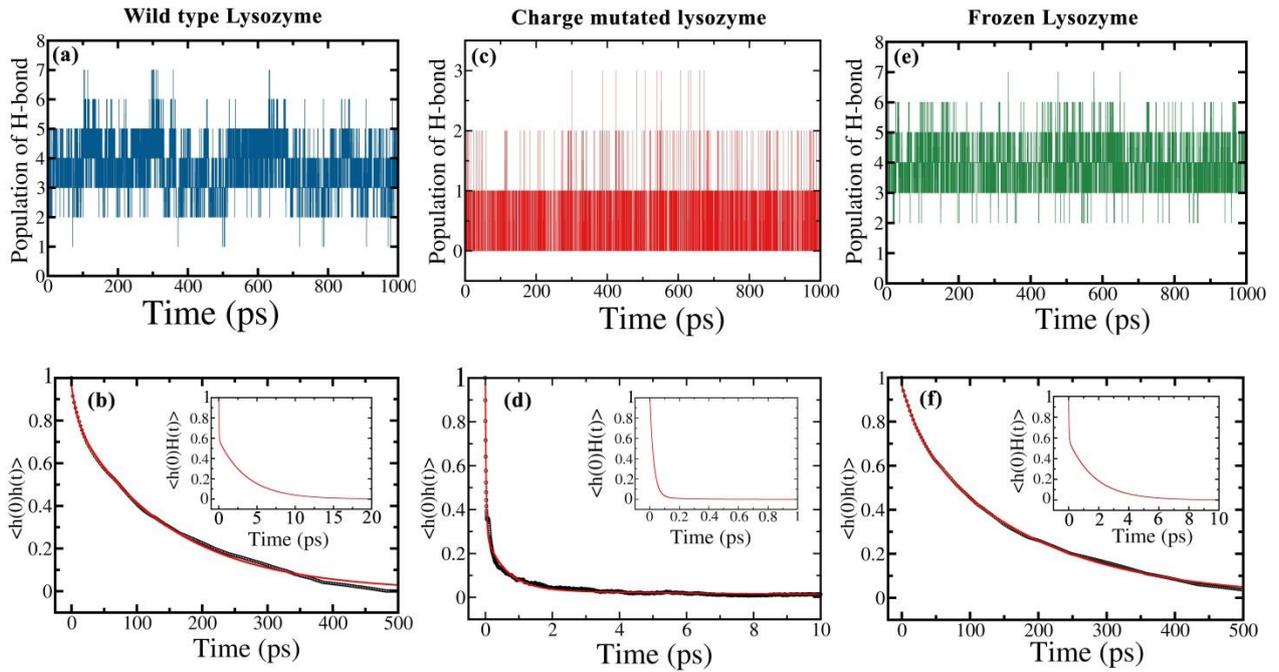

**Figure 8.** Plots representing the population and hydrogen bond lifetimes around W123 of Lysozyme. **(a), (c)** and **(e)** represent the population of H-bond between the surrounding charged residues and water molecules. **(b), (d)** and **(f)** correspond to the relaxation of intermittent H-bond time correlation function. And the continuous H-bond time correlations are given in the insets. Reduction of the charges drastically reduces the H-bond population and relaxation timescales, but freezing does not.

We observe that reduction of the charges destroys the H-bond network between amino acid side chains and water to a huge extent (**Figure 8c**). However, freezing doesn't affect the time constants for intermittent H-bond function to that extent, but it lowers the timescales of continuous H-bond TCF. The latter is more informative on the lifetime of H-bonds. It also becomes clear from this study that long lived H-bond network is one of the governing factors of slow solvation.

**Table 4. Multi-exponential fitting parameters for intermittent and continuous H-bond time correlation functions that are constructed using Heaviside step function.**

| Probe | Correlation function | Surroundings | $a_1$ | $\tau_1 (ps)$ | $a_2$ | $\tau_2 (ps)$ | $a_3$ | $\tau_3 (ps)$ | $<\tau>(ps)$ |
|---|---|---|---|---|---|---|---|---|---|
| W123 | $\dfrac{<h(0)h(t)>}{<h(0)>}$ | Wild Type, Charged | 0.04 | 0.05 | 0.15 | 9.12 | 0.81 | 149.9 | 122.8 |
| | | Mutated, Charges removed | 0.71 | 0.03 | 0.26 | 0.61 | 0.03 | 12.5 | 0.55 |
| | | Wild type, Frozen | 0.03 | 0.09 | 0.18 | 26.7 | 0.79 | 181.1 | 147.8 |
| W123 | $\dfrac{<h(0)H(t)>}{<h(0)>}$ | Wild Type, Charged | 0.42 | 0.04 | 0.58 | 3.71 | --- | --- | 2.17 |
| | | Mutated, Charges removed | 0.97 | 0.03 | 0.03 | 0.19 | --- | --- | 0.03 |
| | | Wild type, Frozen | 0.34 | 0.03 | 0.66 | 1.75 | --- | --- | 1.2 |

## IV. ORIGIN OF BOTH SLOW AND ULTRAFAST SOLVATION

It has been observed earlier that there is approximately a 60-70% sub 100 fs ultrafast component in the aqueous solvation of natural probes. This presence is ubiquitous. In our analysis as we have divided the response arising from water into PHL and outer layer responses, we have found that a significant contribution to the ultrafast component also arise from bulk part (**Figure 6**). This fact can be rationalised by a TDDFT based molecular hydrodynamic theory description developed by Chandra and Bagchi[45,60,61] back in the 1980s. The ultrafast dynamical components (libration, h-bond excitations and single particle rotation) of a dipolar liquid couple to the solvent polarisation to give rise to ultrafast solvation. We can write the expression of free-energy, assuming harmonic approximation, for polarisation fluctuation as in **Eq.[8]**.

$$F(\{P_L(q)\}) = \frac{(2\pi)^3}{2V} \int dq K_L(q) P_L^2(q) \quad [8]$$

Here, $P_L(q)$ corresponds to the longitudinal component of polarisation fluctuation and $K_L(q)$ represents the wavenumber dependent force constant of $P_L(q)$. V is the volume. We can further write $K_L(q)$ in terms of wave number dependent dielectric function of the dipolar medium as **Eq.[9]**.

$$K_L(q) = \frac{2}{(2\pi)^2} \frac{\varepsilon_L(q)}{\varepsilon_L(q) - 1} \quad [9]$$

$\varepsilon_L(q)$ assumes a large value at $q \rightarrow 0$ limit. Hence, $K_L(q)$ converges to $2/(2\pi)^2$. At large q, $\varepsilon_L(q)$ becomes unity and $K_L(q)$ diverges. Solvation energy doesn't derive much contribution from large $q$ limit. Hynes *et. al.*[62,63] earlier showed, with respect to the solvation of an ion, that the energy relaxation can be modelled as a relaxation in a harmonic polarisation potential. The curvature of such a well can be determined by $K_L(q)$. At small wavenumbers $K_L(q)$ is large and curvature is steep, hence it results in a faster relaxation.[48] The ultrafast component is essentially a long wavelength ($q=0$) phenomena which is probed by the bulk solvent. Whereas, the same theory predicts relaxation is slow at large wavenumbers (smaller distances) which is probed by the PHL.

## V. TIME DEPENDENT DENSITY FUNCTIONAL THEORY ANALYSIS

In order to develop a quantitative understanding of the important roles played by several self and cross energy correlation terms in the protein hydration layer dynamics, particularly those between amino acid side chain and water, we need a microscopic theoretical formulation. The only phenomenological theory at hand is the Nandi-Bagchi[6,64] theory but that cannot explicitly describe the different terms involved. Such a microscopic formulation could be hard to achieve given the complexity of the system involved.

Here we follow the previous works of Chandra and Bagchi[41,61] and find that time dependent density functional theory (TDDFT) can capture many of the interesting aspects at a semi-quantitative level.

Solvation energy measured by a probe (as given by **Eq.[2]**) is a function of position ($r$), orientation ($\Omega$) and time dependent polarisation which itself is expressed in terms of position, orientation and time dependent density field ($\delta\rho(r,t)$) at that position.

$$\delta E_{solv}(r,\Omega,t) = -\int dr' d\Omega' C(r-r',\Omega,\Omega')\delta\rho(r',\Omega',t). \quad [10]$$

Here, the primed variables are for the surrounding dipoles around a dipolar probe and '$C$' is the direct correlation function. This leads to the solvation time correlation function as the following,

$$\langle \delta E_{solv}(0)\delta E_{solv}(t)\rangle = \frac{1}{4\pi V}\int dr d\Omega \langle \delta E_{solv}(r,\Omega,0)\delta E_{solv}(r,\Omega,t)\rangle \quad [11]$$

Where, $V$ is the total volume of the system in which a free probe can reside in. But in our case the natural probes are bound to the protein that only allows a partial volume element to be considered. Now we break the integral in **Eq.[11]** into two terms with respect to two major contributions arising due to (i) amino acid side chain atoms and (ii) surrounding water molecules. This gives rise to **Eq.[12]**.

$$\delta E_{solv}(r,\Omega,t) = -\int dr' d\Omega' C_{p-s}(r-r',\Omega,\Omega')\delta\rho_s(r',\Omega',t) \\ -\int dr' d\Omega' C_{p-sc}(r-r',\Omega,\Omega')\delta\rho_{sc}(r',\Omega',t) \quad [12]$$

Here, '$p$', '$sc$' and '$s$' stand for probe, side-chain and solvent respectively. For further simplicity, we assume that the probe bound to the biomolecule as well as the surrounding amino acid side chain atoms are discrete charges instead of dipoles. So, we can drop the orientation dependence of side-chain and probe and rewrite **Eq.[12]** as,

$$\delta E_{solv}(r,t) = -\int dr' d\Omega' C_{p-s}(r-r',\Omega')\delta\rho_s(r',\Omega',t) \\ -\int dr' C_{p-sc}(r-r')\delta\rho_{sc}(r',t) \quad [13]$$

Now, we fix the intermolecular frame along $(r-r')$ as z-axis and convert the integrals into integrals in Fourier space in order to apply density functional theory on solvation energy.

For further progress, the first term on the right hand side of **Eq. [13]** requires expansion of the direct correlation function and the density fluctuation term in terms of spherical harmonics, as explained in the molecular hydrodynamic theory. In the following we suppress the sum over (*l,m*) coefficients, for simplicity. We evaluate the energy time correlation function (**Eq.[13]**). This treatment gives rise to three terms in the solvation time correlation function, two self and one cross term. Hence, the expression becomes the following,

$$\langle \delta E_{solv}(0)\delta E_{solv}(t)\rangle = \frac{1}{(2\pi)^3}\Big[\int d\mathbf{k}\, C_{p-s}^2(\mathbf{k})S_s(\mathbf{k},t) \\ + \int d\mathbf{k}\, C_{p-sc}^2(\mathbf{k})S_{sc}(\mathbf{k},t) \\ + \int d\mathbf{k}\, C_{p-s}(\mathbf{k})C_{p-sc}(\mathbf{k})\langle \delta\rho_s(-\mathbf{k},0)\delta\rho_{sc}(\mathbf{k},t)\rangle \Big] \quad [14]$$

First two terms are self-terms and the third one corresponds to the cross term. It is clear that the first two has no dependency on the sign of the direct correlation function C($\mathbf{k}$) as they come in squares. But the third term carries a cross multiplication and its sign of the product becomes important, which is negative in some of our cases (**Figure 4**).

Furthermore, the dynamic structure factor, denoted by $S_s(\mathbf{k},t)$, of solvent decays fast in the long wave number limit, and hence responsible for the fast decaying component similar to that of ion-solvation in bulk liquid. On the other hand, the $S_{sc}(\mathbf{k},t)$ term is decay slowly and results in the slow components in the decay. The partial structure factor in the third term i.e. $\langle \delta\rho_s(-\mathbf{k},0)\delta\rho_{sc}(\mathbf{k},t)\rangle$ may be either slow or fast depending on the location of the probe. Therefore, by freezing the protein motions we remove the slow contribution completely. This is the reason for the accelerated dynamics is observed and also reported by Singer *et al.*[12,50]

# VI. CONCLUSIONS

In the following we summarise the key results of the present study.

(i) We have used molecular dynamics simulations to carry out studies where we mutate the nature of charge distribution of the amino acid side chains that surround the probe molecule. Since the neighbouring charged side-chain residues play an important role in slowing down the solvation of tryptophan, this study allows us to isolate the role played by other distant residues and also water molecules.

(ii) We also observe the presence of dynamical coupling between amino acid side chain motions and water. *The slowest component in solvation dynamics with time constant above or around 100 ps can only survive if the amino acid side chain motions are present, irrespective of charged/uncharged surroundings.*

(iii) Even when the protein is frozen, there exists a slow component in the range of 20-80 ps with approximately 5-15% relative amplitude. We attribute this residual slowness to the slow water molecules inside PHL. We surmise that this was the component observed by Zewail and co-workers.[7,42]

(iv) The ultrafast component arises from the bulk solvent that contributes via the long wavelength polarization modes and almost entirely responsible for the large amplitude of the sub-100 fs ultrafast solvation. Whereas, the slowness arises due to the water molecules close to the probe (i.e., short wavelength modes). This rather paradoxical result can be rationalized with the molecular hydrodynamic theory of Chandra and Bagchi[41,60,61] as discussed in **Sec IV**.

In a previous work, we have reported that the distributions of rotational and translational relaxation times of individual water molecules change from Gaussian in the bulk to log-normal in the PHL[29]. The ubiquitous long tail in these distributions can be related to the slow component observed in solvation dynamics.

We provide a schematic diagram (**Figure 9**) to elaborate the values and the sources of these timescales that arise due to various contributions, highlighting the contributions from water, as observed in the present as well as many earlier studies.

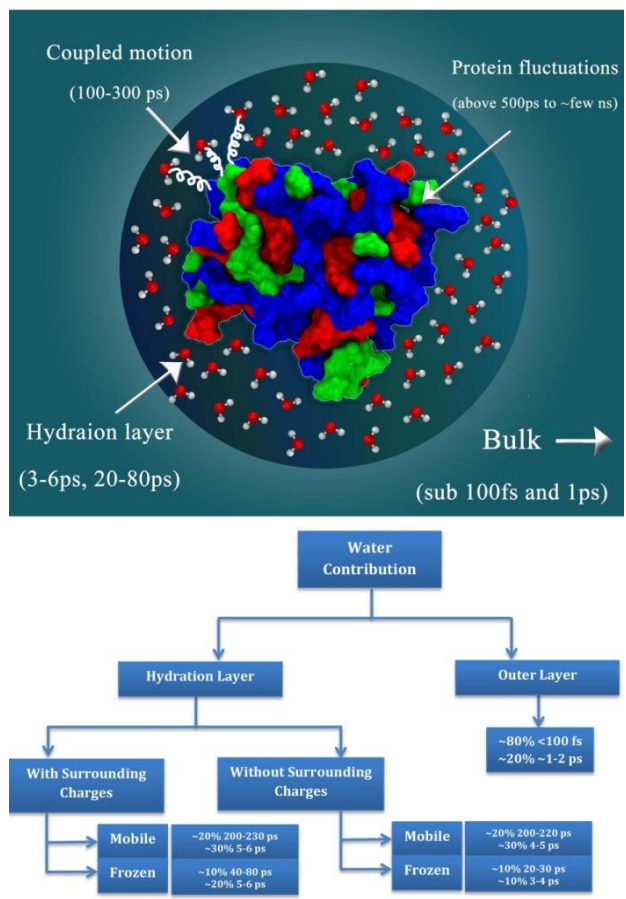

**Figure 9.** A schematic representation of the various sources of slow and ultrafast solvation time scales along with their approximate values solvation of a natural probe at protein hydration layer. Contribution from each part is individually normalised while calculating the % contributions. (*The figure on the top serves as a schematic diagram and strictly not to scale*)

One should bear in mind that there is a considerable heterogeneity across the protein structures and dynamics (*Supporting Information* **S2**) which affects solvation dynamics to a great extent. So no general conclusion is fair to be drawn. However, we have attempted to draw certain general conclusions with the help of time dependent density functional theory. We have suggested that the origin of slow decay mediated by the amino acid side chain and water coupled motions can be described in terms of a partial (or, cross) dynamic structure factor between side-chain atoms and water. The challenge is to precisely calculate the cross-dynamic structure factor between side-chain atoms and water. This serves as an interesting quantity to be calculated from both simulations and theory. The cross correlation functions hold the key to the observed slow/fast dynamics.

Thus, although the search for general features has been elusive, we nevertheless know why PHL in lysozyme behave differently from that in myoglobin by studying the structure of the exposed amino acid side chains. We know that water molecules around charged residues form rigid hydrogen bond structure that slows dynamics by more than one order of magnitude.[6] The flow chart given in **Figure 9** summarizes the generality present in protein hydration layer, and captured through the present study. We also conclude that the reliability of solvation dynamics as a probe to measure the dynamical response depends critically on the site specificity which can be regarded as both strength and weakness of solvation experiments. There is a major scope of extensive theoretical study that should consider proteins of many kinds and also experimental investigations in order to extract certain general conclusions.

## 6. SUPPLEMETARY MATERIALS

We provide Supporting Information to further elaborate, support and to ensure the reproducibility of our results. In **Section S1**, we provide the mutated charges of the neighbourhood of natural probes with proper atomistic pictures of the side-chains. In **Section S2**, we furnish some salient features of the proteins which may affect the dynamics of PHL and in turn affect solvation timescales.

## 7. ACKNOWLEDGEMENTS

We gratefully acknowledge the partial support by grants from DST, India. BB acknowledges support from Sir J.C Bose fellowship. S. Mondal thanks UGC, India for financial support and fellowship. S. Mukherjee thanks INSPIRE, India for providing fellowship.

**Appendix: System and Simulation Details**

Atomistic molecular dynamics simulations have been performed using GROMACS (v-5.0.7)[65]. We construct the systems so that it matches the experimental concentration (~2-3 mM). We have used Optimized Parameters for Liquid Simulation-all atom (OPLS-AA) force field[66] and extended point charge (SPC/E) water model. Periodic boundary conditions were implemented using cubic boxes of sides 93 Å with 26,338 water molecules for Lysozyme (PDB ID: 1AKI)[67]; 94 Å with 26,236 water molecules for sperm whale myoglobin (PDB ID: 3E5O) and 93 Å with 25,652 water molecules for sweet protein Monellin (PDB ID: 3MON). The total system was energy minimised using steepest descend followed by conjugate gradient method. Then the system is subjected to the simulated annealing[68] to heat it up from 300K to 320K and again cool it down from 320K to 300K; in order to get the system out of a local minima (if any). The solvent is equilibrated for 10 ns at constant temperature (300 K) and pressure (1 bar) (NPT) by restraining the positions of the protein atoms followed by NPT equilibration for another 10 ns without position restrain. The final production runs are carried out at a constant temperature (T=300 K) (NVT) for 55 ns. The equations of motions are integrated using leap-frog integrator with an MD time step of 1.0 fs. For analysis, the trajectories were recorded for the last 50 ns with 10 fs resolution. All reported data are averaged over three MD trajectories starting from entirely different configuration of the system.

We have used Nóse-Hoover thermostat[69] and Parrinello-Rahman barostat[70] to keep the temperature and pressure constant respectively. The cut-off radius for neighbour searching and non-bonded interactions has been taken to be 10 Å and all the bonds are constrained using the LINCS[71] algorithm. For the calculation of electrostatic interactions, we have used Particle Mesh Ewald (PME)[72] with FFT grid spacing of 1.6 Å.

# SUPPLEMENTARY MATERIAL

*Abstract*

***In order to find out the microscopic origin of slow component in the dynamics of solvation of a natural probe located, inside protein hydration layer, we mutate some charges around a certain probe. In the main text we discussed the key results in terms of radial distribution function and solvation time correlation function. Here, in the supplementary materials, we provide the details of charge mutations to ensure the reproducibility of our calculations.***

**S1. Details of charge mutation:**

The charges in the wild type (WT) of proteins are acquired from OPLS-AA (Optimized Potential for Liquid Simulation-all atom) force field and those are mutated (MUT) to provide the probes with a net-non polar environment.

| Residue | Atom name | Charges (WT) | Charges (MUT) |
|---|---|---|---|
| Arginine | N | -0.5 | -0.5 |
| | H | 0.3 | 0.3 |
| | CA | 0.14 | 0.14 |
| | HA | 0.06 | 0.06 |
| | CB | -0.12 | -0.12 |
| | HB1 | 0.06 | 0.06 |
| | HB2 | 0.06 | 0.06 |
| | CG | -0.05 | -0.167 |
| | HG1 | 0.06 | 0.02 |
| | HG2 | 0.06 | 0.02 |
| | CD | 0.19 | 0.06 |
| | HD1 | 0.06 | 0.02 |
| | HD2 | 0.06 | 0.02 |
| | NE | -0.7 | -0.293 |
| | HE | 0.44 | 0.147 |
| | CZ | 0.64 | 0.213 |
| | NH1 | -0.8 | -0.326 |
| | HH11 | 0.46 | 0.153 |
| | HH12 | 0.46 | 0.153 |
| | NH2 | -0.8 | -0.326 |
| | HH21 | 0.46 | 0.153 |
| | HH22 | 0.46 | 0.153 |
| | C | 0.5 | 0.5 |
| | O | -0.5 | -0.5 |
| Net Charge | --- | **+1.00** | **0.00** |

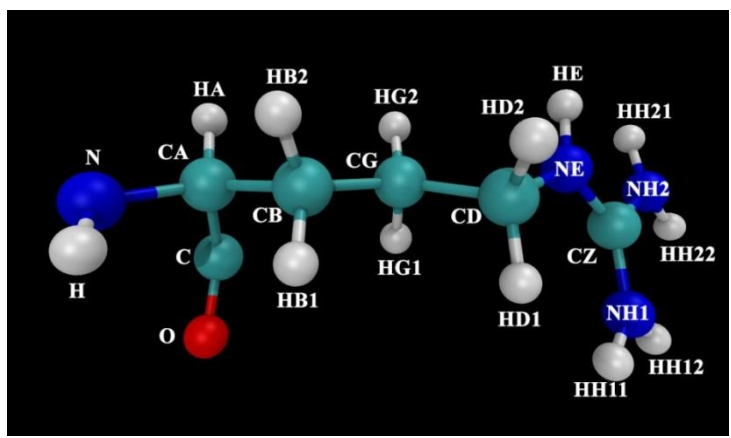

| Residue | Atom name | Charges (WT) | Charges (MUT) |
|---|---|---|---|
| Aspartic acid | N | -0.5 | -0.5 |
| | H | 0.3 | 0.3 |
| | CA | 0.14 | 0.14 |
| | HA | 0.06 | 0.06 |
| | CB | -0.22 | -0.12 |
| | HB1 | 0.06 | 0.06 |
| | HB2 | 0.06 | 0.06 |
| | CG | 0.7 | 0.4 |
| | OD1 | -0.8 | -0.2 |
| | OD2 | -0.8 | -0.2 |
| | C | 0.5 | 0.5 |
| | O | -0.5 | -0.5 |
| Net Charge | --- | **-1.00** | **0.00** |

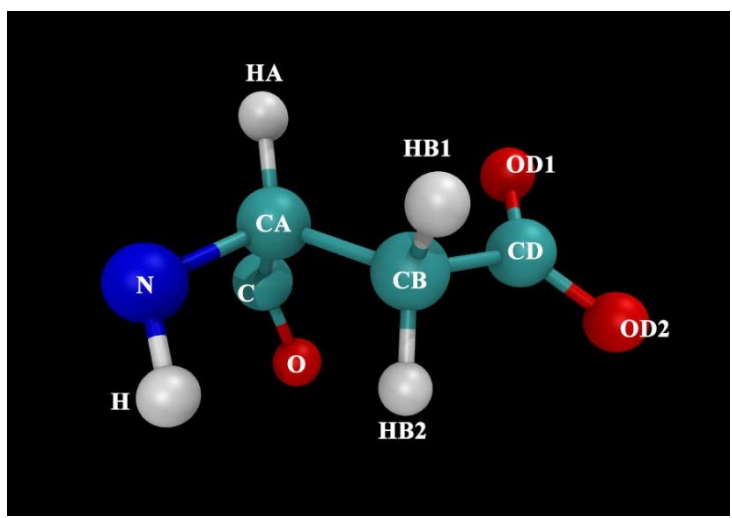

| Residue | Atom name | Charges (WT) | Charges (MUT) |
|---|---|---|---|
| Glutamic Acid | N | -0.5 | -0.5 |
| | H | 0.3 | 0.3 |
| | CA | 0.14 | 0.14 |
| | HA | 0.06 | 0.06 |
| | CB | -0.12 | -0.12 |
| | HB1 | 0.06 | 0.06 |
| | HB2 | 0.06 | 0.06 |
| | CG | -0.22 | -0.12 |
| | HG1 | 0.06 | 0.06 |
| | HG2 | 0.06 | 0.06 |
| | CD | 0.7 | 0.4 |
| | OE1 | -0.8 | -0.2 |
| | OE2 | -0.8 | -0.2 |
| | C | 0.5 | 0.5 |
| | O | -0.5 | -0.5 |
| Net Charge | --- | **-1.00** | **0.00** |

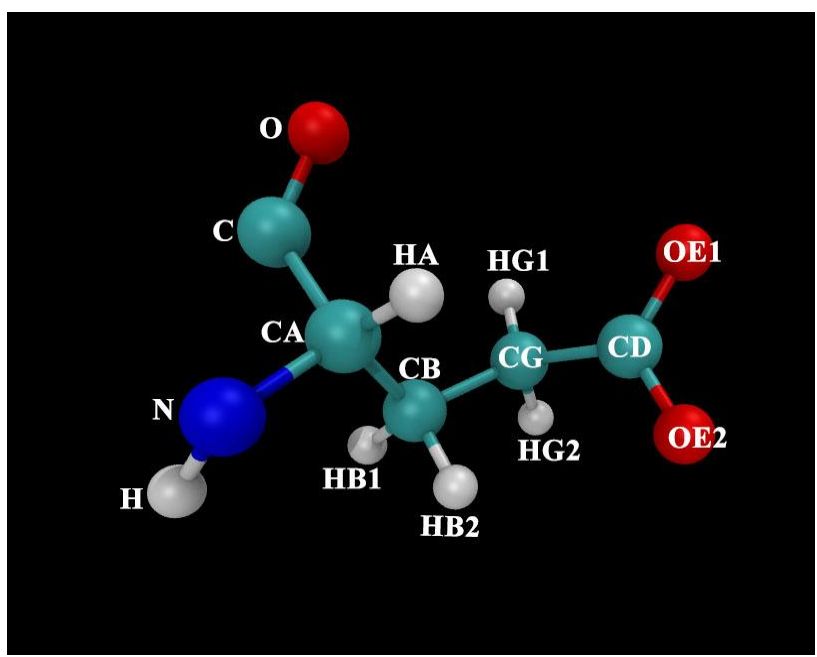

| Residue | Atom name | Charges (WT) | Charges (MUT) |
|---|---|---|---|
| Lysine | N | -0.5 | -0.5 |
| | H | 0.3 | 0.3 |
| | CA | 0.14 | 0.14 |
| | HA | 0.06 | 0.06 |
| | CB | -0.12 | -0.12 |
| | HB1 | 0.06 | 0.06 |
| | HB2 | 0.06 | 0.06 |
| | CG | -0.12 | -0.12 |
| | HG1 | 0.06 | 0.06 |
| | HG2 | 0.06 | 0.06 |
| | CD | -0.12 | -0.12 |
| | HD1 | 0.06 | 0.01 |
| | HD2 | 0.06 | 0.01 |
| | CE | 0.19 | 0.09 |
| | HE1 | 0.06 | 0.01 |
| | HE2 | 0.06 | 0.01 |
| | NZ | -0.3 | -0.10 |
| | HZ1 | 0.33 | 0.03 |
| | HZ2 | 0.33 | 0.03 |
| | HZ3 | 0.33 | 0.03 |
| | C | 0.5 | 0.5 |
| | O | -0.5 | -0.5 |
| Net Charge | --- | **+1.00** | **0.00** |

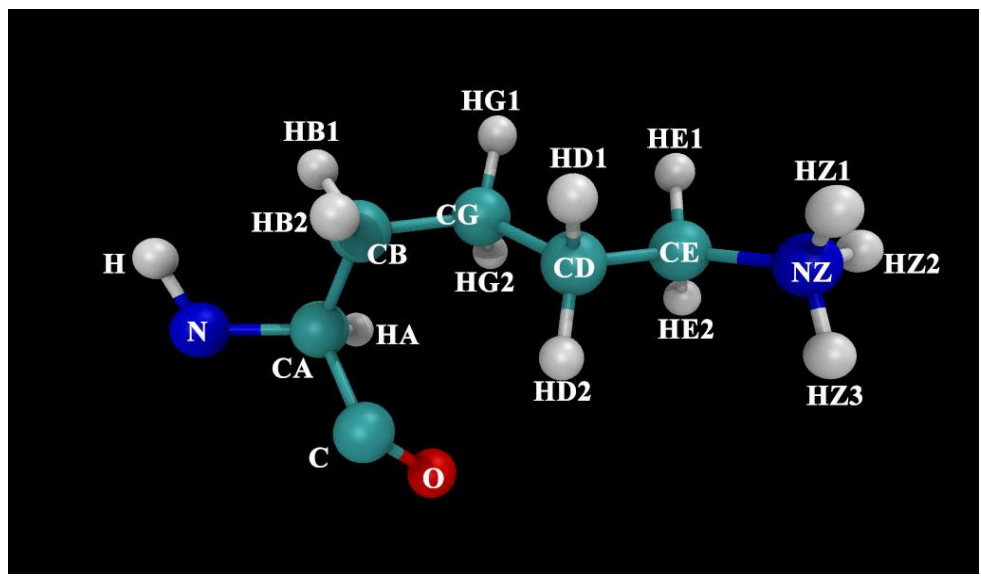

**S2. Some Salient features of Lysozyme, Myoglobin and Monellin**

We have chosen three different proteins to study the solvation dynamics of natural probe tryptophan. Nevertheless, the proteins are unique in their structural and dynamical properties. This heterogeneity provides a differing environment to the probes, which considerably affects the solvation timescales. Below we point out some of the distinct differences.

- Myoglobin is rich in α-helices. There are no β-sheets present. Lysozyme contains both of the secondary structures with an increased population of α-helices. On the contrary, sweet protein Monellin has a greater population of β-sheets.

- There are eleven positively charged arginine residues present in lysozyme. Arginine and lysine are the one of the most hydrophilic amino acids according to certain hydropathy scales. Moreover, hydrophilic solvent accessible surface area (SASA) is ~60% for lysozyme. This polarises the PHL water molecules more than the other two proteins.

- A measure of the structural fluctuation can be obtained in terms of the root mean square deviation (RMSD) trajectory. The RMSD at time 't', in **Fig. S1**, is calculated by superimposing the structure at times 't' and 't-dt', where 'dt' is the data dumping frequency of the MD trajectory. The plot clearly shows that the structural rigidity is quite different for the three protein-water systems. Myoglobin shows the maximum fluctuation. Whereas lysozyme exhibits the least deviation. It tells the increased stability of lysozyme. However, Monellin lies in between.

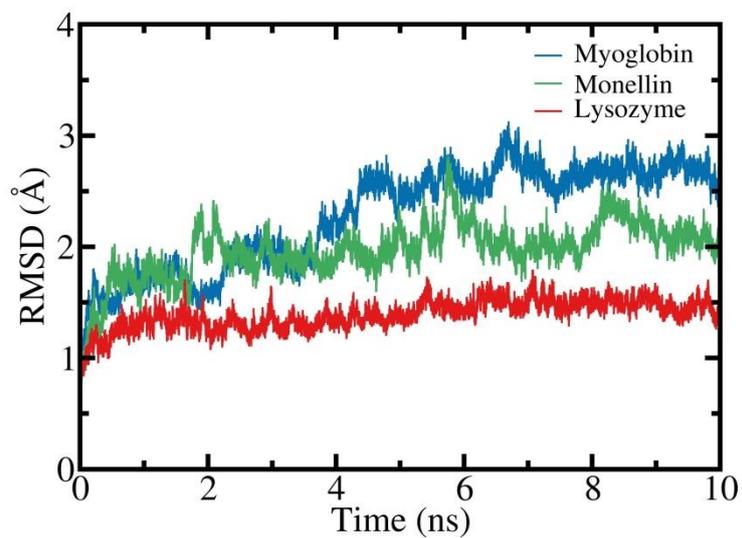

**Figure S1.** Plot of root mean squared deviation (RMSD) of Myoglobin, Lysozyme and Monellin against time with respect to their structure at the previous MD frame. This plots shows the net dynamic heterogeneity among these structures.